\documentclass{PoS}

\usepackage{epsfig,amsmath}

\def\be{\begin{equation}}
\def\ee{\end{equation}}

\title{Correlations in impact-parameter space in saturation models}

\ShortTitle{Correlations in impact-parameter space in saturation models}

\author{\speaker{S. Munier}\\
        Centre de physique th\'eorique, \'Ecole Polytechnique,
CNRS,
91128 Palaiseau, France\\
        E-mail: \email{Stephane.Munier@cpht.polytechnique.fr}}

\abstract{In order to get an estimate of the homogeneity of 
the distribution of matter in a fast hadron, we compute 
the correlation of the saturation
scales between different impact parameters.
We find that these correlations are quite strong: The saturation
scale is nearly uniform in a wide domain around each point in
impact-parameter space. We provide analytical expressions for the
correlations, which are supported by numerical simulations.
Although the numerical calculations are done for specific
saturation models which are obtained from QCD after
drastic simplifications,
we expect our analytical formulas to be correct
for full QCD in asymptotic limits, 
since their derivation requires only
a few general assumptions.
}

\FullConference{XVIII International Workshop on Deep-Inelastic 
Scattering and Related Subjects, DIS 2010\\
April 19-23, 2010\\
Firenze, Italy}

\begin{document}

\section{Introduction}
Phenomenological models for the very high-energy regime of
QCD where saturation effects become important
(see Ref.~\cite{Gelis:2010nm} for a recent review) are usually
built on some parametrization of 
the elastic dipole-hadron scattering amplitude $T(y,r,b)$
which is a function of the rapidity $y$ 
of the scattering,\footnote{%
Throughout our discussion, $y$ is actually the rapidity
multiplied
by the factor $\bar\alpha=\alpha_s N_c/\pi$.
} 
of the size $r$ of the dipole, 
and of the impact parameter $b$.
This amplitude is then related to the observables
through appropriate convolutions with the wave functions
that describe the incoming objects.
In the simplest of these models, due to Golec-Biernat 
and W\"usthoff \cite{GolecBiernat:1998js},
the dipole amplitude 
is assumed to have the form
\be
T(r,y,b)=1-e^{-r^2 Q_s^2(y,b)/4},
\ee
where the momentum scale $Q_s$, called the
saturation momentum, is parametrized as
\be
Q_s^2(y,b)=1\ \mbox{GeV}^2\times
\theta(R-b)e^{\lambda(y-y_0)}.
\ee
The constants $R$, $\lambda$ and $y_0$ 
are determined from a fit to the
inclusive
deep-inelastic scattering data.
The spatial distribution of matter in the plane
transverse to the collision axis
is encoded in the $b$-dependence of the saturation momentum.
The $\theta$-function 
used by Golec-Biernat and W\"usthoff
is sometimes changed 
to a smoother distribution
in such a way that the model be also able to describe
semi-inclusive
diffractive data.
In any case, the fluctuations between different points
in transverse space are completely
neglected in all these models.
Note that this may not be a problem for standard 
phenomenology since most of the observables in
deep-inelastic scattering
probe one single point in impact-parameter space
in each event.
But clearly,
independently of phenomenology,
we would like to understand better
how the matter is distributed 
in a fast hadron.

We shall first explain why fluctuations 
of the parton densities are expected
between different impact parameters,
then we shall provide a heuristical discussion of
the form of these fluctuations, for which 
we have been able to
write a parameter-free 
formula valid
in some asymptotic limit.

\section{Picture of a fast hadron/nucleus}

\begin{figure}
\begin{center}
\includegraphics[width=.9\textwidth]{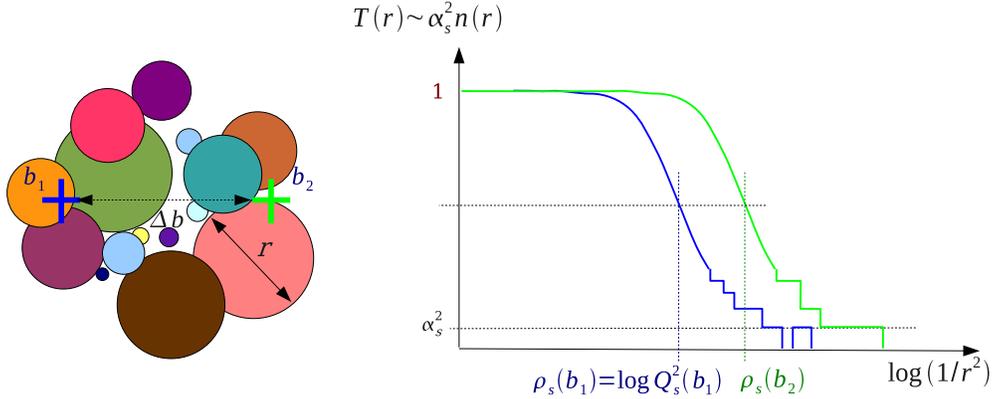}
\end{center}
\caption{\label{fig:1}Schematic picture 
of a fast-moving hadron (left; the colored disks represent partons) 
and scattering amplitude
as a function of the size $r$ of the probing dipole
at two impact parameters $b_1$, $b_2$
for a fixed rapidity $y$ (right).}
\end{figure}

Let us consider a fast hadron or nucleus probed by a color
dipole of size $r$ (which may be seen as a component of
a virtual photon of virtuality $Q\sim 1/r$)
at very high rapidity $y$.
We go to
a frame in which the probing dipole is almost at rest
and we require that the impact parameter be some fixed $b$ 
(see Fig.~\ref{fig:1}).
The scattering probability $T$ is roughly proportional to the local
density $n$ of partons in the corresponding phase-space cell:
$T(r,y,b)\simeq\alpha_s^2 n(r,y,b)$. 
It proves useful to see $T$ as a 
probability of interaction 
between the dipole and
a fixed configuration of partons: 
The
physical amplitude measured in experiments is then
$T$ averaged over
events (i.e. over the partonic configurations; 
see Ref.~\cite{Munier:2009pc} for a review).
If the rapidity is high enough, we know that at each $b$,
$T$ has the shape
of a front connecting 1 (black or
saturated regime) for $r\gg 1/Q_s(y,b)$ to 0 (color transparent or
dilute regime) for $r\ll 1/Q_s(y,b)$. The saturation
momentum $Q_s(y,b)$
determines the transition.
It grows exponentially
with $y$, which means that the position of the
wave front moves linearly 
along the axis $\log(1/r^2)$ when the rapidity increases.
It was first conjectured \cite{Munier:2009pc} and then 
checked numerically \cite{Munier:2008cg}
that to a good approximation,
at each point in impact-parameter space,
the only variable relevant to the
evolution of the amplitude $T$ 
with the rapidity $y$
is $\log(1/r^2)$.

In the regions in which
$T\sim\alpha_s^2$, few partons are probed, 
hence the further evolution of $T$
with the rapidity $y$ is stochastic.
If $T\gg \alpha_s^2$ instead, 
many partons populate that phase-space region,
and the evolution of $T$ is
of deterministic nature:
A mean-field approximation of
the QCD evolution can be taken.
Thus for values of $r$ of the order of the 
inverse saturation momentum,
$T$ has the shape of a smooth (deterministic) curve
traveling towards smaller values of $r$.
However, because of the fluctuations in the tail of the front,
$Q_s$ is a stochastic variable for the rapidity evolution.
Fluctuations in the dilute region
of phase space
propagate towards the dense region and
affect the saturation momentum 
typically after 
an additional evolution over
$\Delta y\sim \log^2(1/\alpha_s^2)$
units of rapidity. 
They result in a random
diffusion of $\rho_s\equiv\log Q_s$
of
variance $\langle \rho_s^2\rangle_c\sim D y$,
where $D$ can be computed from QCD
\cite{Munier:2009pc}.

These fluctuations determine a dispersion of $\rho_s$
from event to event.
But stochasticity is also expected to manifest itself 
by differentiating the
points say $b_1$ and $b_2$
in impact-parameter space, creating
a dispersion of $\rho_s$ in the transverse plane.
In order to characterize these fluctuations,
we shall now compute the correlator
$\sigma_{12}^2\equiv \langle(\rho_s(b_1)-\rho_s(b_2))^2\rangle$
at fixed $y$.

\section{How correlations may occur: heuristic discussion and analytical 
formulas}

\begin{figure}
\begin{center}
\includegraphics[width=.5\textwidth]{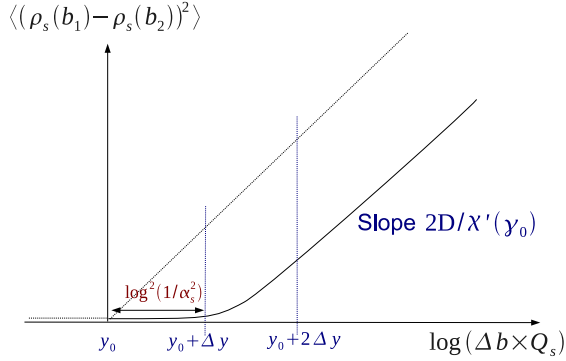}
\end{center}
\caption{\label{fig:2}
Sketch of the correlations as a function of
the logarithm of the distance $\Delta b\equiv|b_1-b_2|$ 
in impact-parameter space scaled
by $1/Q_s$. For $\Delta bQ_s>1$, the points
$b_1$ and $b_2$
are statistically independent.
The dotted line represents what one
would naively expect
if fluctuations affected 
the saturation scale as soon as $\Delta bQ_s>1$.
($D$ is the diffusion coefficient of $\rho_s$
for a single front, namely $D\sim \langle\rho_s^2\rangle_c/y$).
The continuous line takes into account
the delay induced by the
propagation of the fluctuations, which results
in an effective persistence of the correlations.
}
\end{figure}

Let us examine how correlations 
between two points in transverse space
$b_1$ and $b_2$ may build up.
We define $\Delta b=|b_2-b_1|$.
If $\Delta b<1/Q_s$ ($Q_s$ is the
saturation momentum at either $b_1$ or $b_2$), 
then obviously
$Q_s(b_1)=Q_s(b_2)$ and $\sigma_{12}^2=0$.
If $\Delta b>1/Q_s$ instead,
then
the evolution around the impact parameter $b_1$ 
can influence the evolution around $b_2$
only if a parton at $b_1$ splits into a parton
of size of the order of $\Delta b$.
But the saturation of the density of partons 
of sizes larger than $1/Q_s$
disfavors such splittings.
Hence we may think that
the evolutions decouple
as soon as the saturation radius $1/Q_s$ becomes 
smaller than $\Delta b$.
Assume that this happens at rapidity $y_0$: Then
for $y<y_0$, $\sigma_{12}^2=0$, and for $y>y_0$, 
$\sigma_{12}^2\simeq\langle\rho_1^2\rangle_c
+\langle\rho_2^2\rangle_c\sim 2D(y-y_0)$.
One may fix the rapidity
$y$ and vary the distance $\Delta b$ instead:
Then $\sigma_{12}^2\sim 2D \log(\Delta b Q_s)/\chi^{\prime}(\gamma_0)$ 
for $\log(\Delta b Q_s)>0$
(see the dotted line in Fig.~\ref{fig:2}), which suggests
that the characteristic distance scale
for the correlations
in the transverse plane is $1/Q_s$. $\chi(\gamma_0)$ is
a particular eigenvalue of the BFKL kernel $\chi$,
and $\chi^\prime(\gamma_0)$ the asymptotic rate of change of 
$\langle \rho_s\rangle$ with the rapidity \cite{Munier:2009pc}.

However, this is not yet the correct answer.
Indeed, as recalled before, for fluctuations to be able to
differentiate $b_1$ and $b_2$, $\Delta y\sim\log^2(1/\alpha_s^2)$ 
extra
units
of rapidity are needed after the rapidity $y_0$ at which 
$\Delta b Q_s(y_0)=1$.
Hence the effective decoupling of the saturation momenta 
is expected later in rapidity, or for larger distances $\Delta b$.
The correlations would persist over
distances 
$\Delta b\sim e^{c\log^2(1/\alpha_s^2)}/Q_s$ (see the full line
in the sketch of Fig.~\ref{fig:2}).

Extending the phenomenological theory for stochastic fronts developed
in Ref.~\cite{Brunet:2005bz},
we are able to fully compute the correlator 
$\sigma_{12}^2\equiv\langle(\rho_s(b)-\rho_s(b+\Delta b))^2\rangle$. 
One way of writing the result is \cite{Mueller:2010fi}
\be
\sigma_{12}^2=
\frac{2\pi^2}{3\gamma_0^2\log(1/\alpha_s^2)}
\int_{\exp\left\{
-\frac{\pi^2\gamma_0^2\chi^{\prime\prime}(\gamma_0)
\left[\log(1/\alpha_s^2)/\gamma_0+\log(\Delta b Q_s)\right]
}{2\chi^\prime(\gamma_0)\log^2(1/\alpha_s^2)}
\right\}}^1
\frac{dq}{q}\left[-\partial_q\vartheta_4(0|q)\right],
\label{res}
\ee
where $\vartheta_4$ is a particular Jacobi theta function.
The interesting limiting behaviors read
\be
\sigma_{12}^2
\sim
\begin{cases}
\frac{2\pi^4\chi^{\prime\prime}(\gamma_0)\log(\Delta b Q_s)}
{3\chi^{\prime}(\gamma_0)
\log^3(1/\alpha_s^2)}
&\mbox{\small for $\log(\Delta b Q_s)\gg\log^2(1/\alpha_s^2) $}\\
\frac{4}{3\gamma_0^3}
\sqrt{\frac{2\pi^3\chi^\prime(\gamma_0)}{\chi^{\prime\prime}(\gamma_0)
\log(\Delta b Q_s)}}
\exp\left(
-\frac{\chi^\prime(\gamma_0)\log^2(1/\alpha_s^2)}
{2\gamma_0^2\chi^{\prime\prime}(\gamma_0)\log(\Delta b Q_s)}
\right)
&\mbox{\small
for $\log(1/\alpha_s^2)\ll\log(\Delta b Q_s)
\ll \log^2(1/\alpha_s^2)$.}
\end{cases}
\ee
Comparing the expression
of $\sigma_{12}^2$ in the large $\Delta b$ 
limit to
the variance $Dy$ of $\rho_s$, we find that
$\sigma_{12}^2$ is actually equal to
$2 D\log(\Delta b Q_s)/\chi^\prime(\gamma_0)$ 
for large $\log(\Delta b Q_s)$.
From the second limiting expression, it is obvious
that $\sigma_{12}^2$ is close to zero for 
$\log(\Delta b Q_s)\ll\log^2(1/\alpha_s^2)$.

\begin{figure}
\begin{center}
\includegraphics[width=.7\textwidth]{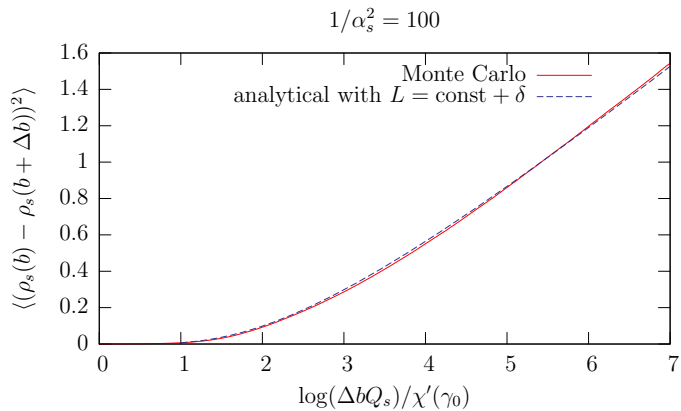}
\end{center}
\label{fig:3}
\caption{%
Comparison of a numerical Monte Carlo simulation
and our analytical formula.
The constant in the parameter $L$ 
(see Ref.~\cite{Mueller:2010fi} for the definitions 
of $L$ and $\delta$),
which should be equal to
$\log(1/\alpha_s^2)/\gamma_0$ for very small $\alpha_s$, 
has been shifted by a phenomenological
constant. Once this is done, we get 
a very good agreement
between the two calculations.
}
\end{figure}

In order to check these expressions,
we performed numerical simulations of models
which possess the main characteristics of the QCD evolution
while being simple enough to allow for
robust
Monte Carlo simulations
(see Ref.~\cite{Mueller:2010fi} for details).
We found a perfect matching with the parameter-free 
analytical result~(\ref{res})
in the limit $\log(1/\alpha_s^2)\gg 1$.
For larger and more realistic values of $\alpha_s$, 
the persistence of the correlations
is still seen in the numerical simulations, 
but some parameters
should be modified
in the analytical expressions
and tuned to account for 
our lack of understanding of subleading corrections
important
for finite $\log(1/\alpha_s^2)$. We show such a calculation
for $\alpha_s=0.1$ in Fig.~\ref{fig:3}, 
compared to a variant
of Eq.~(\ref{res}).

\section{Conclusion and outlook}

The main result of our work is that 
the characteristic distance
of the correlations
in the transverse plane is not $1/Q_s$ as one would naively expect,
but rather $\exp\left[c\log^2(1/\alpha_s^2)\right]/Q_s$ 
($c$ being a known constant),
which is parametrically much larger than $1/Q_s$.
Our results are valid for large $\log(1/\alpha_s^2)$, and
for distances 
$\Delta b$ much
smaller than the typical confinement scale 
$1/\Lambda_{\mbox{\small QCD}}$.

The goal of our work was to understand the 
fundamentals of the
QCD dynamics in transverse space, without thinking
{\it a priori} of any application
to phenomenology.
Let us however note that recently, 
a diffractive 
deep-inelastic scattering 
observable was proposed that would
directly probe the correlations which we have computed
\cite{Marquet:2010cf}.
(A calculation of these correlations
in the framework of the B-JIMWLK formalism
\cite{Gelis:2010nm}
which {\it a priori} neglects the fluctuations
discussed in this paper
was also performed.)
Also, these correlations may play an important
role in heavy-ion collisions.


\begin{thebibliography}{99}

\bibitem{Gelis:2010nm}
  F.~Gelis, E.~Iancu, J.~Jalilian-Marian and R.~Venugopalan,
  arXiv:1002.0333 [hep-ph].

\bibitem{GolecBiernat:1998js}
K.~J. Golec-Biernat, M.~W\"usthoff, 
Phys. Rev. D59 (1999) 014017.

\bibitem{Munier:2009pc}
S. Munier,
  {}Phys.\ Rept.\  { 473} (2009) 1.

\bibitem{Munier:2008cg}
S.~Munier, G.~P. Salam, G.~Soyez, 
Phys. Rev. D78 (2008) 054009.

\bibitem{Brunet:2005bz}
  E.~Brunet, B.~Derrida, A.~H.~Mueller and S.~Munier,
  Phys.\ Rev.\  E{73} (2006) 056126.

\bibitem{Mueller:2010fi}
{}A.~H.~Mueller and S.~Munier,
Phys.\ Rev.\  D{81} (2010) 105014.

\bibitem{Marquet:2010cf}
  C.~Marquet and H.~Weigert,
  arXiv:1003.0813 [hep-ph].



\end{thebibliography}
\end{document}